\documentclass{appolb}
\usepackage{epsfig}

\newcommand{\beqn}{\begin{eqnarray}}
\newcommand{\eeqn}{\end{eqnarray}}
\newcommand{\be}{\begin{equation}}
\newcommand{\ee}{\end{equation}}

\newcommand{\ba}{\begin{array}{c}}
\newcommand{\bat}{\begin{array}{cc}}
\newcommand{\ea}{\end{array}}

\newcommand{\bi}{\begin{itemize}}
\newcommand{\ei}{\end{itemize}}

\newcommand{\Frac}[2]{\frac{\displaystyle #1}{\displaystyle #2}}

\newcommand{\mL}{\mathcal{L}}

\newcommand{\mF}{\mathcal{F}}

\newcommand{\cO}{{\cal O}}

\newcommand{\ket}{\,\rangle}
\newcommand{\bra}{\langle \,}

\begin{document}


\title{$1/N_C$ Expansion of QCD Amplitudes%
\thanks{Talk presented at {\it Effective theories
of colours and flavours: from EURODAPHNE to EURIDICE},  24-27 August
2006, Kazimierz (Poland).   This work  has been supported by
EU~RTN~Contract~CT2002-0311 and  China National Natural Science
Foundation under grant number 10575002 and  10421503.} }

\author{ J.J. Sanz-Cillero
\address{Department of Physics, Peking University, Beijing
100871, P.R. of China  \\
e-mail: cillero@th.phy.pku.edu.cn} }

\maketitle

\vspace*{-1cm}
\begin{abstract}
This talk comments the main features of a hadronic description of
QCD in the limit of large number of colours. We derive a quantum
field theory for mesons based on chiral symmetry and a perturbative
expansion in $1/N_C$. Some large--$N_C$  and next-to-leading order
results are reviewed.
\end{abstract}

\vspace*{-0.5cm}
\section{Introduction}

The success of large--$N_C$ determinations based on 't Hooft's limit
of QCD~\cite{NC} has led to an extraordinary development of the
field and has naturally raised the question about subleading
effects~\cite{Cata}--\cite{HLS}.
The analysis beyond leading order in $1/N_C$ (LO) is essential to
validate the large--$N_C$ limit.  A formally well defined $1/N_C$
expansion can be easily obtained by implementing the proper $N_C$
scalings of the hadronic couplings and
masses~\cite{NC,VFFrcht,functional,SSPPrcht}. On the other hand,
naively, loops with heavy resonance are expected to produce
corrections of the form $\frac{M_R^2}{16\pi^2 F^2}$, order $1/N_C$
but numerically large. However, previous phenomenological analyses
have shown that the next-to-leading order contributions (NLO) remain
under control~\cite{VFFrcht,SSPPrcht}. Rewriting the resonance
parameters in terms of widths ($\Gamma_R$) an masses ($M_R$), one
observes that the $1/N_C$ expansion is actually an expansion around
the narrow-width limit. The corrections to the large--$N_C$
amplitudes are suppressed by $\Gamma_R/M_R\sim 1/N_C$. However, it
is not yet clear how broad-width  states like the $\sigma$ meson fit
in this pattern~\cite{Hanqing,Pelaez}.

Since the QCD action is chirally invariant, one needs to construct a
chiral theory for resonances (R$\chi$T) that preserves the symmetry.
This feature, common to several phenomenological
lagrangians~\cite{HLS,rcht-Ecker,rcht-op6,spin1fields,rcht-Donoghue},
ensures the recovery of Chiral Perturbation Theory ($\chi$PT)~\cite{chpt}
at low energies even at the loop level.
Likewise, the validity of the $1/N_C$ expansion at all energies  allows  to
match QCD at  short distances, where it is described by the  operator product expansion
(OPE)~\cite{OPE}.

\section{Large--$N_C$, next-to-leading order and resummations}

In the large--$N_C$ limit, QCD contains an infinite tower of
hadronic states, the resonances ($R$) and the Goldstones from the
spontaneous chiral symmetry breaking ($\phi$). The Green-functions
are provided by the tree-level exchanges~\cite{NC}. Other
observables like the form-factors are derived from them and they are
also given by the tree-level diagrams. Their absorptive contribution
is a series of delta functions , so the amplitudes are determined by
the masses (position of the real poles) and the corresponding
couplings (residues).

At NLO in $1/N_C$, the perturbative   amplitudes contain two-meson
absorptive cuts together with single and double real poles coming,
respectively,  from diagrams with one and  two tree--level
propagators~\cite{VFFrcht,SSPPrcht}.
Pure perturbation theory, i.e.,
without resummation, is valid when no intermediate particle is  near
its mass-shell.

However, the perturbative expansion breaks down  in the
neighbourhood of the resonance poles at any finite order in $1/N_C$
and a Dyson-Schwinger summation is required~\cite{Dyson}. This
shifts the real resonance poles into unphysical Riemann sheets. The
particles gain a finite width and the amplitude becomes finite.

In the past, the attention has been focused either on the
large-$N_C$ limit or  on resummed descriptions. However, the
previous step to any resummation is the perturbative calculation and
only a few examples of it exist by the
moment~\cite{Cata}--\cite{HLS}.
By-passing this intermediate point may lead to strongly model
dependent resummations and, therefore, incorrect determinations.

\vspace*{-0.3cm}
\section{Resonance Chiral Theory}

\subsection{Leading order in $1/N_C$}

In general,  one is forced to work within a minimal hadronical
approximation with a finite number of states (MHA)~\cite{MHA}. This
is an acceptable approximation in the case of amplitudes that are
chiral-order parameter, provided that we include a minimal number of
light states, enough to reproduce the short-distance QCD power
behaviour.

Since we work within a large--$N_C$ framework,  the particles are
classified in $U(n_f)$ multiplets~\cite{anomaly}. The Goldstones
from the spontaneous chiral symmetry breaking
$\phi=(\pi,K,\eta_8,\eta_0)$ are incorporated through covariant
tensors ${\mathcal G}(\phi)$~\cite{rcht-Ecker,rcht-op6,chpt}. The
lightest $1^{--},\, 1^{++},\, 0^{++},\, 0^{--}$ resonance fields are
included, being the spin--1 mesons represented through antisymmetric
tensors $R^{\mu\nu}$~\cite{rcht-Ecker,spin1fields}. The last
ingredient of R$\chi$T  relies on the assumption that operators with
a large number of derivatives are forbidden and only $\cO(p^2)$
chiral tensors are to be considered. The addition of higher power
operators is expected to lead to a wrong growing behaviour of the
Green-functions at large momenta. These ingredients yield the
general lagrangian,
\begin{equation}
\mL_{\rm R\chi T}=\mL_{\rm G}(\phi)\, +\, \sum_{\rm R} \mL(R,\phi)\,
+\, \sum_{\rm R,R'} \mL(R,R',\phi)\, +...
\end{equation}
The operators with just Goldstones are given  by $\chi$PT at
$\cO(p^2)$~\cite{chpt}:
\begin{equation}
\mL_{\rm G}(\phi)\, =\, \Frac{F^2}{4}\bra u_\mu u^\mu \, +\,
\chi_+\ket\, .
\end{equation}
The operators linear in the resonance fields were derived in Ref.~\cite{rcht-Ecker}:
\begin{equation}
\sum_{\rm R}\mL(R,\phi)\, =\, \Frac{F_V}{2\sqrt{2}}\bra
V_{\mu\nu}f_+^{\mu\nu}\ket \, +\, \Frac{i G_V}{\sqrt{2}}\bra
V_{\mu\nu}u^\mu u^\nu\ket \, +\,  c_d \bra S \chi_+\ket \, + ...
\end{equation}
The  analysis of
three-point functions and form-factors have requires the
introduction of operator with two and three resonance
fields~\cite{SSPPrcht,rcht-op6,3point}.

In order to make the theory dual to QCD, it must be  matched at the
regions where it is calculable, this is, at low and high energies.
The recovery of the low energy QCD structure, described by $\chi$PT,
is trivial once chiral symmetry has been properly incorporated. On
the other hand, R$\chi$T  must reproduce the OPE at short
distances~\cite{OPE}. For instance, the matching of the $V-A$
correlator yields the well known Weinberg sum-rules and establishes
a relation between resonance parameters at LO in
$1/N_C$~\cite{PI:02}.

\subsection{Next-to-leading order}

The one-loop diagrams give place to NLO contributions . They produce
ultraviolet (UV) divergences that require new operators with NLO
couplings in order to be renormalised. Many of these operators can
be actually removed through the use of the  equations of
motion~\cite{VFFrcht,functional}. Furthermore, it has been proved
that some matrix elements do not need local $\chi$PT operators to
fulfill the renormalisation~\cite{vanishing}. We will refer here to
the example of the correlator  $\Pi(t)\equiv \Pi_{_{\rm
SS}}(t)-\Pi_{_{\rm PP}}(t)$~\cite{SSPPrcht}.

\begin{center}
\begin{figure}
\vspace*{-0.7cm} \center{\includegraphics[width=7cm,clip]{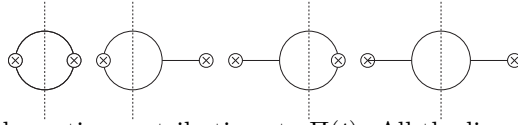}}
\vspace*{-0.9cm} \caption{\small{One-loop absorptive contributions
to $\Pi(t)$. All the lines stand for tree-level meson propagators in
our perturbative calculation. }} \label{fig.diagrams}
\end{figure}
\end{center}

The first step is to examine the absorptive part of the amplitude by
means of the optical theorem.  The contribution from a two-particle
intermediate state $\rm M_1M_2$ to the spectral function, shown in
Fig.(\ref{fig.diagrams}), is in general proportional to some squared
form factors~\cite{SSPPrcht}:
\begin{equation}
\mbox{Im}\Pi(t)_{_{\rm M_1M_2}} \propto \left|\mF_{_{\rm
M_1M_2}}(t)\right|^2\, .
\end{equation}
The vanishing of $\mF_{\pi\pi}(t)$ at infinite
momentum~\cite{Brodsky} makes Im$\Pi(t)_{\pi\pi}\to 0$ when
$t\to\infty$. Demanding  a similar vanishing behaviour for each
separate absorptive contribution Im$\Pi(t)_{_{\rm M_1M_2}}$  leads
to a series of constraints for the form-factors at LO in
$1/N_C$~\cite{SSPPrcht}.

After the preliminary large--$N_C$ analysis of the absorptive
subdiagrams, one is ready to aboard the renormalisation of the
one-particle-irreducible vertex-functions entering in our
amplitude~\cite{VFFrcht,functional}. Imposing the proper UV
asymptotic conditions in the absorptive part of $\Pi(t)$ leads to
the absence of new UV divergent structure~\cite{vanishing}. No new
operators are needed for the renormalisation, just a $1/N_C$ shift
of the parameters existing at LO.

An alternative way to calculate   NLO
amplitudes~\cite{Cata,VFFrcht,HLS} is the use of dispersion
relations~\cite{SSPPrcht}. In our case, it is possible to write down
an unsubtracted dispersive integral for $\Pi(t)$.
%
%
The two-meson cuts contribute to
the correlator with a finite part, $\Delta \Pi(t)_{_{\rm M_1M_2}}$,
and  $\Pi(t)$ depends now on the renormalised resonance  masses and
couplings:
\begin{eqnarray}
\Pi(t) = 2 B_0^2 \left[ \Frac{8 c_m^{r\, 2}}{M_S^{r\, 2}-t} -\Frac{8
d_m^{r\, 2}}{M_P^{r\, 2}-t} +\Frac{F^2}{t}+
\displaystyle{\sum_{\rm M_1M_2}}\Delta \Pi(t)_{_{\rm M_1M_2}}\right] \, .
\end{eqnarray}
The NLO  contribution $\Delta\Pi(t)_{_{\rm M_1M_2}}$
only depends on Im$\Pi(t)_{_{\rm M_1M_2}}$~\cite{SSPPrcht}
and, at high energies,
behaves like
\begin{equation}
\Delta \Pi(t) =  \frac{F^2}{t}\, \delta_{\rm NLO}^{(1)} \,+ \,
\frac{F^2 M_S^2}{t^2}\, \left(\delta_{\rm NLO}^{(2)} +
\widetilde{\delta}_{\rm NLO}^{(2)}\ln{\frac{-t}{M_S^2}}\right)\,\,
\, + \, \, \cO\left(\Frac{1}{t^3}\right)\, .
\end{equation}

The matching of the one-loop R$\chi$T correlator to the OPE yields a
NLO generalisation of  the first and second  Weinberg
sum-rules~\cite{SSPPrcht}:
\begin{eqnarray} 8 c_m^{r\, 2}  - 8 d_m^{r \, 2} -
F^2\, (1+ \delta_{\rm NLO}^{(1)}) &=& 0 \, ,
\nonumber  \\
- 8 c_m^{r\, 2} M_S^{r\, 2} + 8 d_m^{r \, 2} M_P^{r\, 2}+ F^2
M_S^2\, \delta_{\rm NLO}^{(2)} &=&\,  -8\, \widetilde{\delta} \, ,
\label{eq.D4const}
\end{eqnarray}
where $\widetilde{\delta}\equiv 3\pi \alpha_s F^4/4$ turns out to be
numerically negligible~\cite{PI:02}. The matching is completed by
demanding that the $\frac{1}{t^2}\ln{(-t/M_S^2)}$ term also vanishes,
this is, $\widetilde{\delta}_{\rm NLO}^{(2)}=0$. These relations fix
the renormalised resonance couplings $c_m^r$, $d_m^r$ in terms of
the renormalised resonance masses $M_R^r$~\cite{SSPPrcht}.

One-loop R$\chi$T  reproduces at low energies
the one--loop $\chi$PT structure,  generating the proper running
for the chiral couplings. Thanks to this, it is possible to provide
predictions for the renormalised  $L_i^r(\mu)$ at any $\mu$
in terms of the renormalised R$\chi$T parameters.
In our example, the short-distance matching of the form-factors at
large--$N_C$ and the correlator  at NLO fixes  the chiral coupling
$L_8^r(\mu)$ in terms of the renormalised masses
$M_R^r$, yielding the prediction~\cite{SSPPrcht}:
\begin{equation}
L_8^r(\mu_0)= (0.6\pm 0.4)\cdot 10^{-3} \, ,  \qquad\qquad \mbox{for}\quad \mu_0=770\mbox{ MeV}\, .
\end{equation}
The main error, also present at LO, comes from  the scalar and
pseudoscalar masses. The uncertainty on the saturation scale is
completely removed. One must keep in mind that any large--$N_C$
estimate of the LECs contains an inherent theoretical error due to
the NLO running from the loops.  There is no particular saturation
scale for all the $\chi$PT couplings. This uncertainty can be  only
removed by taking the calculation up to the one-loop level.

\section{Open questions}

Although it is possible to extract some information about the
couplings of highly excited mesons ($M_R\gg \Lambda_{\rm
QCD}$)~\cite{alphasPI}--\cite{Regge},
one still needs to specify the structure of the spectrum at high
energies. It can be solved in some models~\cite{Regge}--\cite{Son}
but, in general, the QCD spectrum is badly known in the ultraviolet.
This forces to work under a MHA~\cite{MHA},introducing
uncertainties~\cite{alphasPI,L8-Peris} that are  reflected in some
problems in the short-distance matching of three-point Green
functions~\cite{Prades}. An improved way to perform the matching
would be desirable. In addition to making MHA a complete and
self-consistent description, it would allow the exploration of
Green-functions that are not order parameter and  are actually
dominated by the high part of the infinite series of
resonances~\cite{alphasPI,Regge}.

A last standing problem is the equivalence between large--$N_C$
lagrangians. The spin--1 mesons can be described through different
formulations~\cite{HLS,spin1fields,rcht-Donoghue,Kampf}. However,
the equivalence between representations has been only  proven at
$\cO(p^4)$~\cite{spin1fields,Pallante} and higher orders have not
been explored. Likewise, a general proof forbidding operators of
order higher than $\cO(p^2)$ in the lagrangian is still missing.
Nevertheless, the slow but firm advances in the field are
establishing solid foundations where to base the $1/N_C$ hadronic
phenomenology.



\begin{thebibliography}{90}





\bibitem{NC}
  G.~'t Hooft,
  {\it Nucl. Phys.} B {\bf 72} (1974) 461; \\
%
  E.~Witten,
  {\it Nucl. Phys.} B {\bf 160} (1979) 57.



\bibitem{Cata}
  O.~Cat\`a and S.~Peris,
  {\it Phys. Rev.} D {\bf 65} (2002) 056014.



\bibitem{VFFrcht}
    I. Rosell {\it et al.},
    {\it JHEP}  {\bf 0408} (2004) 042.


\bibitem{functional}
 I.~Rosell {\it et al.},
 {\it JHEP} {\bf 0512} (2005) 020.


\bibitem{SSPPrcht}
    I. Rosell {\it et al.},
    hep-ph/0610290;
%
%
    J.J. Sanz-Cillero,  hep-ph/0610304.

\bibitem{vanishing}
    J. Portol\'es {\it et al.},
    hep-ph/0611375.


\bibitem{HLS}
%
%
    M. Harada and K. Yamawaki,
    {\it Phys. Lett. } B {\bf 297} (1992) 151-158.

\bibitem{Hanqing}
    Z.-H. Guo {\it et al.},
    hep-ph/0610434.

\bibitem{Pelaez}
    J.R. Pelaez,
    {\it Mod. Phys. Lett.} A {\bf 19} (2004) 2879-2894.




\bibitem{rcht-Ecker}
  G.~Ecker {\it et al.},
  {\it Nucl. Phys.} B {\bf 321}(1989) 311.


\bibitem{rcht-op6}
  V.~Cirigliano {\it et al.},
  Nucl.\ Phys.\ B {\bf 753} (2006) 139.


\bibitem{spin1fields}
  G.~Ecker {\it et al.},
  {\it Phys.  Lett.} B {\bf 223} (1989) 425.


\bibitem{rcht-Donoghue}
    J.F. Donoghue {\it et al.},
    {\it Phys. Rev.} D {\bf 39} (1989) 1947.




\bibitem{chpt}
    J. Gasser and H. Leutwyler, {\it Annals Phys.} {\bf 158} (1984)
    142 ;  \\
%
%
    {\it Nucl. Phys.} B {\bf 250} (1985) 465-517.



\bibitem{OPE}
    M.A. Shifman {\it et al.},
    {\it Nucl. Phys.} B {\bf 147} (1979) 385-447;
%
    {\bf 147} (1979) 448-518.






\bibitem{Dyson}
    D. G\'omez-Dumm {\it et al.},
    {\it Phys. Rev.} D {\bf 62} (2000) 054014-1;  \\
%
%
    J.J. Sanz-Cillero and A. Pich, {\it Eur. Phys. J.} C {\bf 27} (2003) 587.




\bibitem{MHA}
  M.~Knecht and E.~de Rafael,
  {\it Phys. Lett.} B {\bf 424} (1998) 335; \\
  M.~Knecht {\it et al.},
  {\it Phys. Lett.} B {\bf 443} (1998) 255;  \\
%
%
    S. Friot {\it et al.},
    {\it JHEP} {\bf 0410} (2004) 043.





\bibitem{anomaly}
    S.R. Coleman and E. Witten,
    {\it Phys. Rev. Lett.} {\bf 45} (1980) 100.




\bibitem{3point}
  B. Moussallam,
  {\it Phys. Rev.} D {\bf 51} (1995) 4939; 
%
    {\it Nucl. Phys.} B {\bf 504} (1997) 381;
%
%
  M.~Knecht and A.~Nyffeler,
    {\it Eur. Phys. J.} C {\bf 21} (2001) 659;  \\
%
%
    P.D.~Ruiz-Femen\'{\i}a {\it et al.},
    {\it JHEP} {\bf 0307} (2003) 003;  \\
%
%
    V.~Cirigliano {\it  et al.},
  Phys.\ Lett.\ B {\bf 596} (2004) 96;    \\
%
%
  V.~Cirigliano {\it et al.},
  JHEP {\bf 0504} (2005) 006.

\bibitem{PI:02}
    For a review of large--$N_C$ constraints see
    A. Pich, hep-ph/0205030.




\bibitem{Brodsky}
    G.P.~Lepage and S.J.~Brodsky, {\it Phys. Lett.} B {\bf 87} (1979) 359;
    {\it Phys. Rev.} D {\bf 22} (1980) 2157;
    {\bf 24} (1981) 1808.






\bibitem{alphasPI}
    J.J. Sanz-Cillero,
    {\it Nucl. Phys.} B {\bf 732} (2006) 136-168.

\bibitem{Cohen}
    T.D. Cohen and E.S. Werbos,  hep-th/0612209.


\bibitem{L8-Peris}
    M. Golterman and S. Peris,
    {\it Phys. Rev.} D {\bf 74} (2006) 096002.



\bibitem{Regge}
    M.A. Shifman,  hep-ph/0009131; \\
%
%
  M.~Golterman and S.~Peris, JHEP {\bf 0101} (2001) 028;  \\
%
%
    S.S. Afonin {\it et al.},
    {\it JHEP}  {\bf 0404} (2004) 039.


\bibitem{NC-1+1}
    G. 't Hooft, {\it Nucl. Phys.} B {\bf 75} (1974) 461; \\
%
%
    J. Mondejar, A. Pineda and J. Rojo,
    {\it JHEP} 0609 (2006) 060.

\bibitem{Son}
    D.T. Son and M.A. Stephanov,
    {\it Phys. Rev.} D {\bf 69} (2004) 065020.


\bibitem{Prades}
    J. Bijnens {\it et al.},
    {\it JHEP} {\bf  0304} (2003) 055.

\bibitem{Kampf}
    K. Kampf, J. Novotny and J. Trnka, hep-ph/0608051.

\bibitem{Pallante}
    J. Bijnens and E. Pallante,
    {\it Mod. Phys. Lett.} A {\bf 11} (1996) 1069-1080.
\end{thebibliography}
\end{document}